\documentclass[aps,prl,twocolumn,showpacs,superscriptaddress]{revtex4-1}

\usepackage{amsmath,braket,amssymb,natbib,graphicx,float,amsthm,xcolor,outlines,mathrsfs,multirow,hhline,bm}

\usepackage{dcolumn}
\usepackage[english]{babel}
\usepackage[utf8x]{inputenc}
\usepackage[T1]{fontenc}
\usepackage{listings}
\usepackage[colorinlistoftodos]{todonotes} 
\usepackage[colorlinks=true, allcolors=blue]{hyperref}
\DeclareMathOperator{\nb}{\bar{{\it n}}}

\begin{document}
\title{Deterministic nonlinear bunching of bosons}
\author{Kingshuk Adhikary}
\altaffiliation{These authors contributed equally to this work.}
\email{kingshuka@imsc.res.in}
\affiliation{Department of Optics, Palack\'{y} University, 17. listopadu 1192/12, 779 00 Olomouc, Czech Republic}
\affiliation{Optics and Quantum Information Group, The Institute of Mathematical Sciences,
HBNI, C. I. T. Campus, Taramani, Chennai 600113, India}
\author{Darren W. Moore}
\altaffiliation{These authors contributed equally to this work.}
\email{darren.moore@upol.cz}
\affiliation{Department of Optics, Palack\'{y} University, 17. listopadu 1192/12, 779 00 Olomouc, Czech Republic}
\author{Radim Filip}
\email{filip@optics.upol.cz}
\affiliation{Department of Optics, Palack\'{y} University, 17. listopadu 1192/12, 779 00 Olomouc, Czech Republic}

\begin{abstract}
The ability of bosonic energy quanta to bunch together in an energy-conserving interaction is a fundamental feature of quantum harmonic oscillators. Linear systems together with measurement allow for the conditional concentration of energy quanta and, subsequently, breeding of the quantum states, but only with an exponentially decreasing success rate. Deterministic, energy-conserving and unconditional bunching however, requires nonlinearity. We investigate which nonlinear energy-conserving interactions deterministically combine bosons into high number states at the same frequency. We show that in order to do so it is advantageous to use nonlinear interactions involving highly saturable systems, such as qubits, as they preserve the hierarchical quantum non-Gaussian features and are also sufficiently robust against pure loss. Nonlinear bunching therefore demonstrates the advantage of a {\it qubit-inside} nonlinearity and opens new directions in the deterministic preparation, processing, and detection of quantum non-Gaussian states.
\end{abstract}

\maketitle

{\it Introduction}---~For quantum harmonic oscillators the energy quanta are fully characterised by a Fock basis of bosonic states~\cite{dirac_principles_2010}. Their bosonic nature means that energy quanta can be brought together in a principally indistinguishable way~\cite{bose_plancks_1924}, even energetically passively. Linear-dynamics methods may bunch bosons, often explored as a method to create higher Fock states~\cite{spagnolo_general_2013,motes_efficient_2016,sturges_quantum_2021}. Bunching interactions that are both energetically passive and linear comprise passing bosons through beamsplitters, with their bunching heralded via detection on some fraction of the output ports. This probabilistic bunching has been widely demonstrated by the Hong-Ou-Mandel effect~\cite{hong_measurement_1987,grosse_measuring_2007,makino_synchronization_2016,marek_direct_2017,ollivier_hong-ou-mandel_2021,lang_correlations_2013,gao_programmable_2018,toyoda_hongoumandel_2015,bouchard_two-photon_2020,alsing_extending_2022,chen_scalable_2023,von_lupke_engineering_2024,yousef_metasurface_2025}. Linear bunching is further explored in boson-sampling problems~\cite{shchesnovich_universality_2016,loredo_boson_2017,spagnolo_non-linear_2023,seron_boson_2023,young_atomic_2024}. However, adding active
feed-forward correction using only displacement or squeezing in linear bunching schemes to remove probabilistic limitation comes at the cost of lower quality bunched states. Recently, different approaches using active interactions like linear amplifiers~\cite{chen_two-particle_2025} and QND interactions~\cite{manukhova_atom-mechanical_2022} can provide new bunching phenomena. Still, they do not allow deterministic bunching to higher Fock states. Nonlinear measurements can improve bunching beyond that possible via linear networks~\cite{francalanci_generalized_2026}.
\begin{figure}[h!]
    \centering
    \includegraphics[width=\columnwidth]{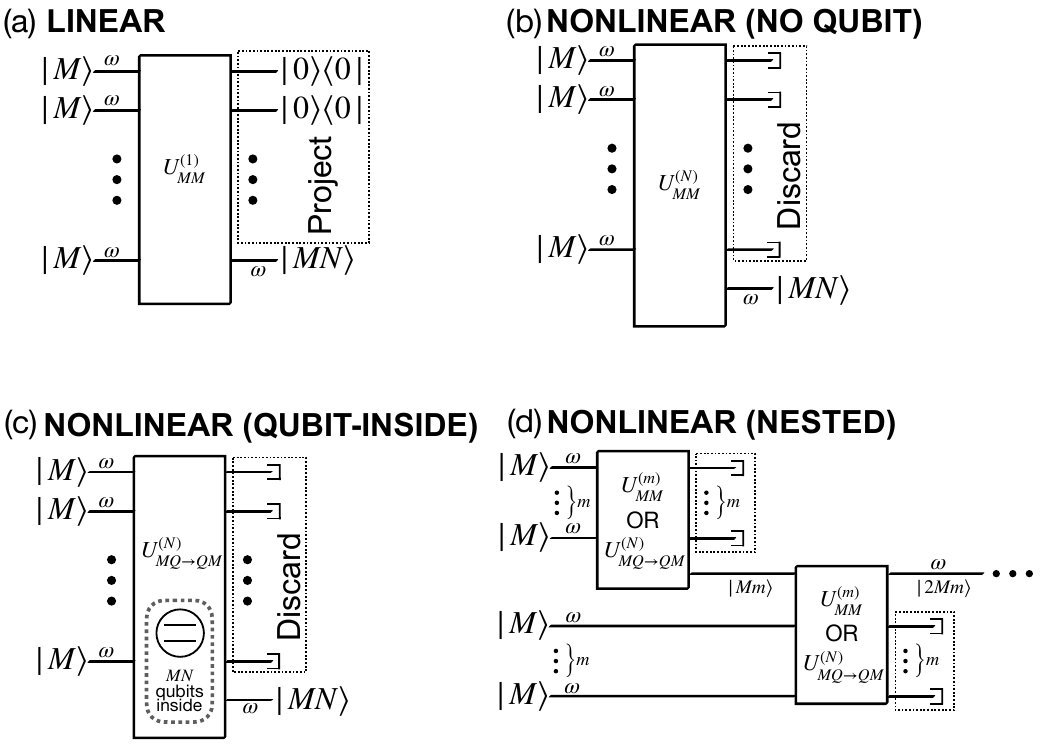}
    \caption{Circuit diagrams of (a) linear and (b) nonlinear energy-conserving (no qubit) interactions $U^{(m)}=e^{-iH^{(m)}t}$ to bunch bosonic energy quanta at frequency $\omega$, with subscripts indicating the interaction type (see Eqs.~(\ref{MW}-\ref{QM})). Linear circuits constructed from interaction (\ref{MW}), with index $m=1$ can never bunch bosons deterministically, but can do so either probabilistically, or conditionally by ground state measurements on almost all modes~\cite{zapletal_experimental_2021}. In contrast, nonlinear energy-conserving (no qubit) circuits constructed from interaction (\ref{MW}), $m=N$ can deterministically bunch bosons into an auxiliary target mode. (c) Deterministic and energy-conserving nonlinear (qubit-inside) scheme in which a nonlinear coupling with a qubit constructed from interactions (\ref{MQ}, \ref{QM}) bunches bosons into an auxiliary target mode. (d) Nesting of lower order nonlinear interactions $U^{(m)}$ to progressively bunch incoming bosons. This fails for fully CV nonlinear (no qubit) interactions but succeeds for qubit-inside nonlinear interactions.}
    \label{Sketch}
\end{figure}

In this Letter we show that by harnessing nonlinear interactions bosons can be deterministically and energetically passively bunched in a linear oscillator, without active transformations, measurements, or heralding, and is robust to input loss, certified by the genuine $n$-photon quantum non-Gaussianity (QNG) hierarchy~\cite{podhora_quantum_2022}. Importantly, in so doing we prove that a strongly saturable nonlinear component is required to efficiently achieve such bunching, which we term a {\it qubit-inside} nonlinearity.

{\it Results}---~Bunching of bosonic energy quanta at the same frequency $\omega$ is described schematically in Fig.~\ref{Sketch}. With linear passive interactions (Fig.~\ref{Sketch}~(a)) ideal bunching can only be achieved conditionally, by projecting all but one output port onto the ground state. This collapses all bosons into a single quantum state. Allowing nonlinear passive CV interactions (Fig.~\ref{Sketch}~(b)) promotes this scheme to a deterministic one albeit at the cost of introducing an additional output mode at the same frequency, prepared in the ground state. An alternative scheme (Fig.~\ref{Sketch}~(c)) involves an intermediate step in which the bosons are redirected to qubits inside via resonant interactions, followed by a nonlinear, passive, and deterministic bunching interaction involving nonlinear coupling of the output mode with the collection of qubits. This qubit-inside nonlinearity has significant implications for bunching. Finally, either scheme can be divided into nested operations with lower-order nonlinearities and fewer modes at each step (Fig.~\ref{Sketch}~(d)), gaining these advantages at the cost of deeper circuits.

Considering energy-conserving unitary processes, we are restricted to Hamiltonian interactions that separately commute with the individual free Hamiltonians and do not exchange energy, or that are in the rotating wave approximation. We discuss three key interactions
\begin{align}
    H^{(m)}_{\text{MM}}&=\left(a_0^\dagger{}^m\prod_{i=1}^ma_i+\text{h.c.}\right)\label{MW}\\
    H^{(m)}_{\text{QM}}&=\left(a_0^\dagger{}^m\prod_{i=1}^m\sigma_-^{(i)}+\text{h.c.}\right)\label{MQ}\\
    H^{(m)}_{\text{MQ}}&=\left(\sum_{i=1}^mb_i\sigma_+^{(i)}+\text{h.c.}\right)\label{QM}
\end{align}
whose subscripts indicate multimode purely CV nonlinear mixing (MM) and  multiqubit interactions (MQ: modes-to-qubits, and QM: qubits-to-modes), representing the simplest case of single bosons arriving in oscillators or via qubit intermediaries respectively (Fig.~\ref{Sketch}, $M=1$). We consider the generalised case $M>1$ later. The $a_i$ and $b_i$ are annihilation operators and $\sigma_\pm$ are qubit raising and lowering operators. Each interaction has an associated index $m$ which specifies the number of initial inputs (oscillators/qubits) carrying a single boson, and possesses an output mode with annihilation operator $a_0$ where the individual bosons are bunched, for a total of $m+1$ subsystems. For Eqs.~(\ref{MW}) and (\ref{MQ}) $2m$ also indicates the order of the nonlinearity. $H^{(1)}_{\text{MM}}$ generates linear bunching ($U_{\text{MM}}^{(1)}$, Fig.~\ref{Sketch}~(a)), while $H^{(m)}_{\text{MM}}$ generates nonlinear bunching (no qubit) ($U^{(N)}_\text{MM}$, Fig.~\ref{Sketch}~(b)) that can be nested ($U^{(m)}_\text{MM}$, Fig.~\ref{Sketch}~(d)). $H^{(m)}_{\text{MQ}}$ describes the collective transfer of single quanta to qubit systems, while $H^{(m)}_{\text{QM}}$ describes the transfer of single quanta from the excited qubits to the output mode. Together these form the nonlinear (qubit-inside) bunching ($U^{(N)}_{\text{MQ$\rightarrow$ QM}}$, Fig.~\ref{Sketch}~(c)). As a consequence of their energy conservation, these interactions preserve the total number of excitations among the subsystems. Note that for $m>1$ the restriction on same frequency can be loosened to resonance conditions of the general form $\omega_0=\sum_i^m\omega_i$. Note the the qubits are restricted to having the same frequency as the delivery modes $b_i$ in Eq.~(\ref{QM}).
\begin{figure}
    \centering
    \includegraphics[width=\columnwidth]{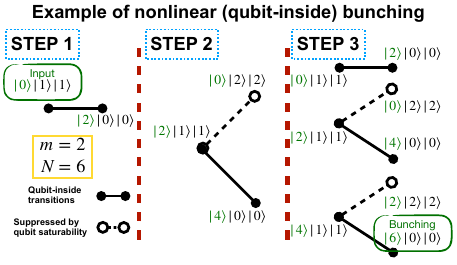}
    \caption{Graphical diagram of the allowed transitions between fixed quanta states for $N=6$ quanta arriving in pairs $m=2$ [see Eqs.~(\ref{MW},~\ref{MQ})] for both fully CV and qubit-inside interactions using the nested circuit, Fig.~\ref{Sketch}~(d). The leftmost ket (green) is the output mode. The suppression of unwanted transitions (dashed lines) by the qubit saturability means there is a clear set of three deterministic steps (solid lines) from the input $\ket{0}\ket{1}\ket{1}$ to the ideally bunched output $\ket{6}\ket{0}\ket{0}$. Without this suppression (no qubit-inside) quanta can leak away from the output mode, making high quality deterministic bunching extremely unlikely. By step 3 the no-qubit-inside strategy evolves on a classical mixture of the three disconnected subgraphs, with different total numbers of quanta, effectively guaranteeing that bunching will not occur.}
    \label{SketchGraphical}
\end{figure}

In order to bunch a collection of single bosons the energetically passive interactions must allow a transition to a fully bunched state and simultaneously block transitions away from this bunched state. If $N$ single quanta in $N$ modes, $\ket{1}^{\otimes N}$, are to bunch together then (\ref{MW}) and (\ref{MQ}) with $m=N$ represent the extreme case where this bunching happens after a single interaction. However when these interactions have nonlinearity order $m<N$ the interaction must be applied sequentially $\frac{N}{m}$ times and provided with $m$ fresh single bosons at each step, as depicted in Fig.~\ref{Sketch}~(d), while carrying over the state of the output mode $0$ between interactions. For Eq.~(\ref{MQ}) these bosons arrive via the intermediate step using Eq.~(\ref{QM}). Naturally this puts restrictions on which $N$ can be targeted, as $m$ must divide the target number of quanta $N$. Finally, the success of the bunching is evaluated by detecting genuine Fock-state attributes using the full rank $n$-photon QNG hierarchy~\cite{chabaud_certification_2021,podhora_quantum_2022} via $P_N=\braket{N|\rho|N}$ where $\rho$ is the final state of the system and $\ket{N}$ is the state of $N$ bosons in mode $a_0$.

As a first result, when the input single boson states are pure we can trivially achieve $P_N=1$ for any $N$ with nonlinearity order $m=N$ with either of the key interactions (\ref{MW}) without any qubits or (\ref{MQ}) with the qubit-inside property. This is because there is a single possible transition, taking the collection of single bosons directly to a bunched state, that is $\ket{0}\ket{1}^{\otimes N}\rightarrow\ket{N}\ket{0}^{\otimes N}$. When $m<N$ and nesting is required this no longer holds and several transitions must be passed through to reach $P_N$. The structure on these transitions induced by the key interactions determines how well the bosons will bunch. 

Let us use this straightforward method already to describe how the nested protocol should ideally work, on an example depicted graphically in Fig.~\ref{SketchGraphical}. Consider the case with $N=6$ quanta, $\ket{1}^{\otimes 6}$, arriving in pairs $m=2$, to be bunched via the nested scheme in Fig.~\ref{Sketch}~(d). In this case the key interactions are run with only $\ket{1}^{\otimes2}$ of the $N=6$ bosons, bunching them ideally into the output mode $a_0$ i.e. $\ket{0}\ket{1}^{\otimes 2}\rightarrow\ket{2}\ket{0}^{\otimes 2}$. Then, the oscillators or qubits $i=1,2$ are replaced with fresh single bosons $\ket{1}^{\otimes2}$ and the interaction is run again, attempting to maximise $P_4$ associated with the transition $\ket{2}\ket{1}^{\otimes 2}\rightarrow\ket{4}\ket{0}^{\otimes 2}$, and then again maximising $P_{N=6}$. 

Now compare strategies based on Eq.~(\ref{MW}) without qubits and Eq.~(\ref{MQ}) with qubit-inside nonlinearity. Fig.~\ref{SketchGraphical} shows that in Step 1, qubit-inside or not, it is possible to ideally bunch the first two bosons. There are only two possible states and the system oscillates between them, allowing a deterministic and full transition from $\ket{0}\ket{1}^{\otimes2}$ to $\ket{2}\ket{0}^{\otimes2}$. At Step 2 the qubit-inside and fully CV systems bifurcate. The CV system under Eq.~(\ref{MW}) has two possible transitions away from $\ket{2}\ket{1}^{\otimes2}$, whereas the saturability of the qubit-inside system under Eq.~(\ref{MQ}) suppresses one of these transitions. This suppression ensures that Step 2 for the qubit-inside interaction still consists of only a single transition, again deterministic and full, from $\ket{2}\ket{1}^{\otimes2}$ to $\ket{4}\ket{0}^{\otimes2}$. In Step 3 the qubit-inside nonlinearity acts the same way as in Steps 1 and 2, guaranteeing ideal bunching $\ket{4}\ket{1}^{\otimes 2}\rightarrow\ket{6}\ket{0}^{\otimes 2}$. For the CV (no qubit) interaction the output at step 2 cannot be prepared in $\ket{4}$ with unit probability $P_4=1$. Instead, after tracing out the input modes the output $a_0$ is left in a classical mixture of the form $p_4(t)\ket{4}\bra{4}+p_2(t)\ket{2}\bra{2}+p_0(t)\ket{0}\ket{0}$. Alongside the final two quanta arriving for Step 3, $\ket{1}^{\otimes2}$, each component of this mixture has different total numbers of quanta and, since this number is conserved, we are completely blocked from ever again achieving a pure state in the output mode. At Step 3, ideal bunching from nesting without use of a qubit has been permanently blocked.

\begin{figure}
    \centering
    \includegraphics[width=0.49\columnwidth]{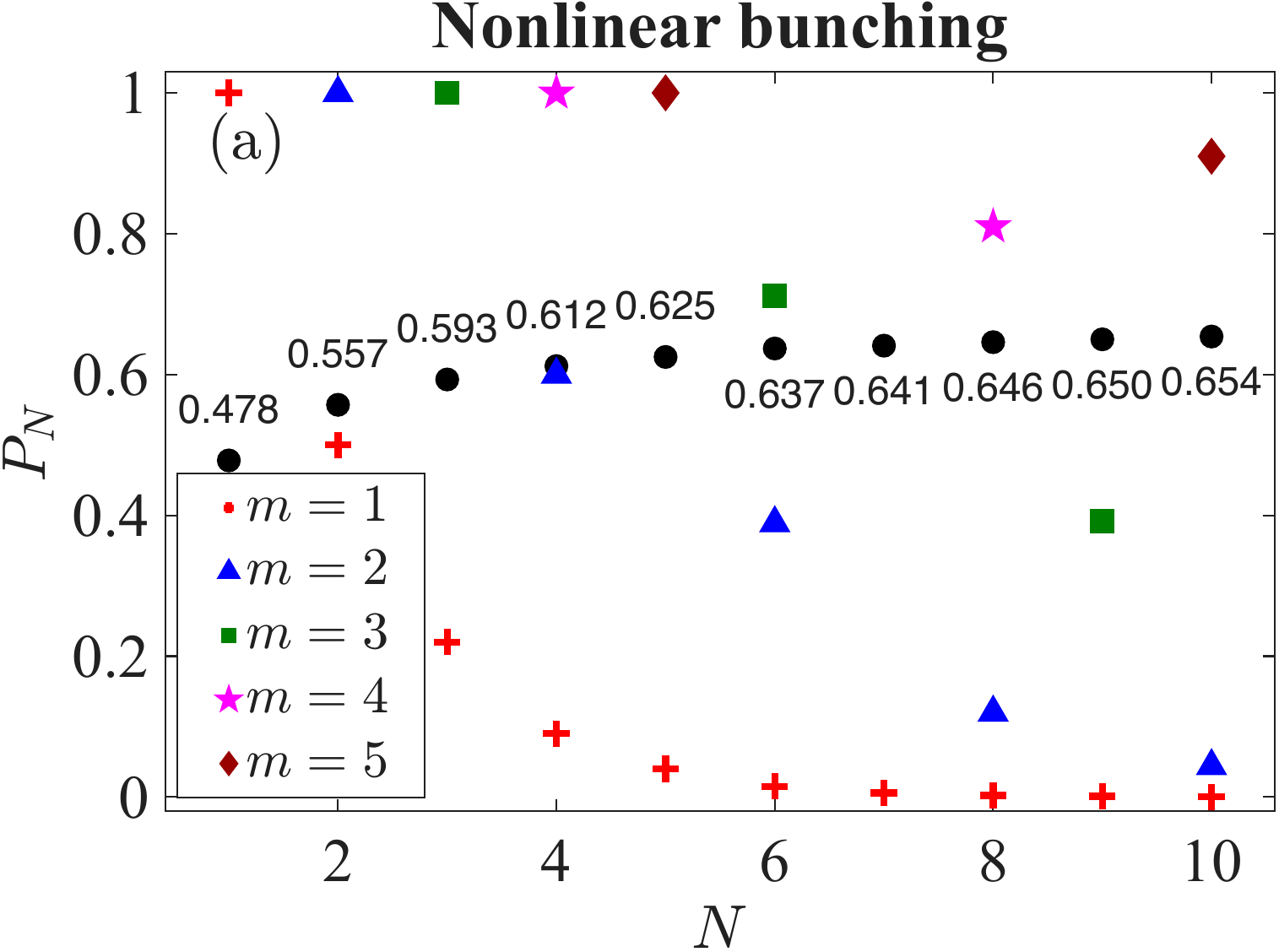}
    \includegraphics[width=0.49\columnwidth]{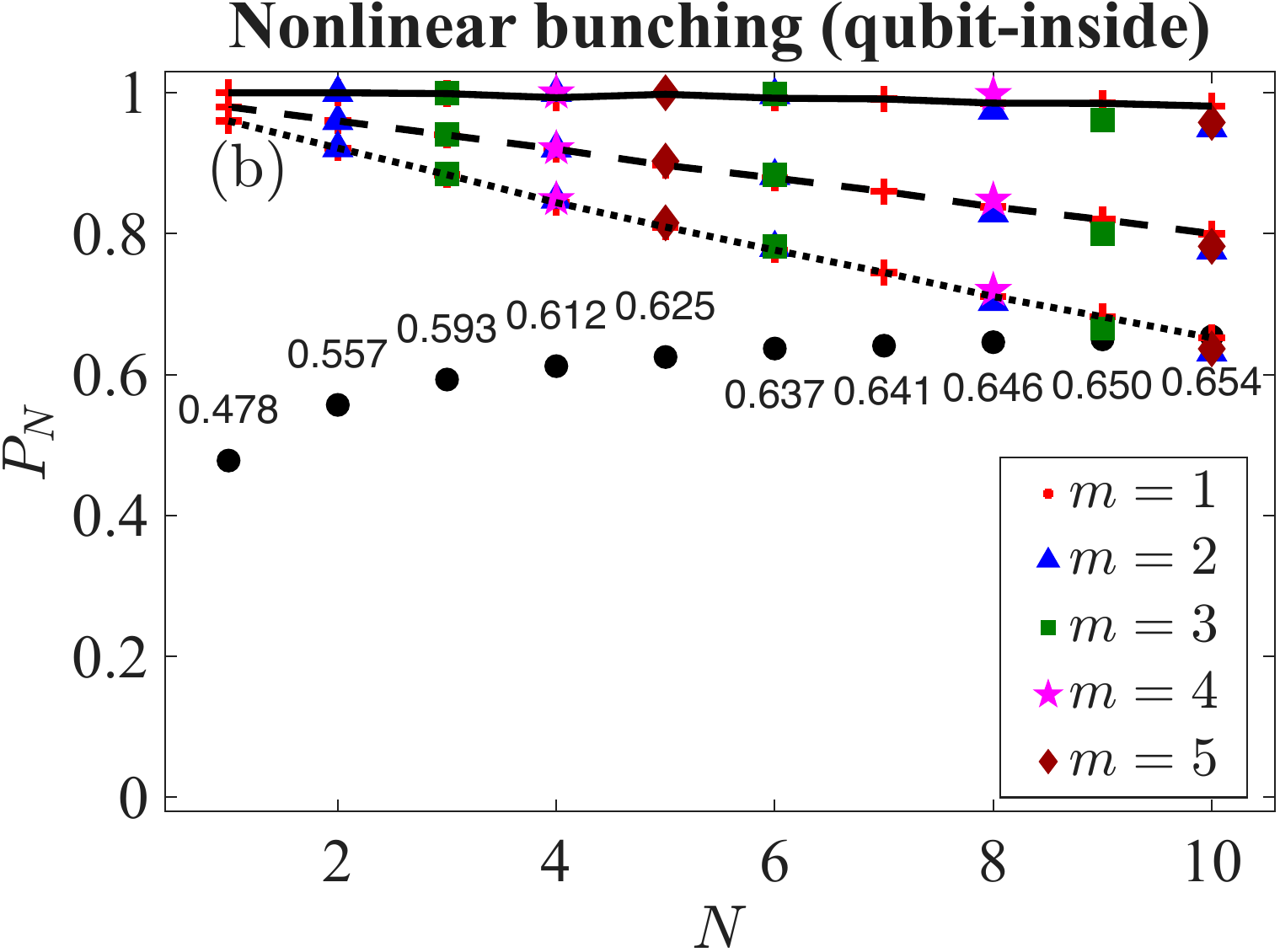}
    \caption{(a) Nonlinear bunching (maximum probability $P_N=\braket{N|\rho|N}$ of the bunched state) using the multimode CV interaction Eq.~(\ref{MW}) for $m\le N$, where $m$ is the number of quanta arriving in a single step. When $m=N$ the bunching is achieved, however an $m=N$ nonlinearity is required in (\ref{MW}). Unfortunately, nested interactions fail to effectively bunch the bosons, with $P_N$ rapidly falling below the $n$-photon genuine QNG thresholds (black points)~\cite{podhora_quantum_2022}. Longer sequences of interactions result in lower probability of ideal bunching, thus $m=1$ is the worst performing protocol which gets worse as $N$ increases. (b) Nonlinear bunching using the qubits-inside nonlinearity based on interactions Eq.~(\ref{MQ}) for $m\le N$ for single bosons with loss $\eta=1$ (solid), 0.98 (dashed) and 0.96 (dotted). The effectiveness of nesting is advantageously independent of $m$ (all markers overlap at fixed $N$) and the sensitivity to loss for a greater target $N$ is linear compared to an exponential decrease without qubits inside.}
    \label{m<Npure}
\end{figure}

Fig.~\ref{m<Npure}~(a) shows that for the no qubit CV nonlinear interaction nesting catastrophically fails to bunch, even for pure single boson states. The greater the difference between nonlinearity order $m$ and the targeted number of quanta at the output $N$, the lower the probability $P_N$ of the targeted bunched state $|N\rangle$ resulting in inconclusive genuine QNG. In striking contrast the qubit-inside protocol, evaluated in Fig.~\ref{m<Npure}~(b), can be nested with pure states to retrieve $P_N=1$ independent of the number of nesting steps indicated by $\frac Nm$. This demonstrates how qubits inside the interactions can reduce the requirements on the increasing number of modes and order of the nonlinearity, both indicated by $m$, for the CV nonlinearity Eq.~(\ref{MW}). The saturability of the qubits produces a barrier for the output mode, so that transitions away from the bunched state are suppressed, as described in Fig.~\ref{SketchGraphical}. In Fig.~\ref{m<Npure}~(b) we further show the nesting for the qubit-inside protocol with initial loss $\eta=0.98,~0.96$ (discussed in detail below). The robustness under loss indicates that the advantage offered by the qubit-inside protocol is not limited to pure initial states. The overlapping points even under loss also hint at the independence of the qubit-inside version from the order $m$ of the nonlinear interaction, a further advantage over the multimode version which we now turn to. 

\begin{figure}
    \centering
    \includegraphics[width=0.49\columnwidth]{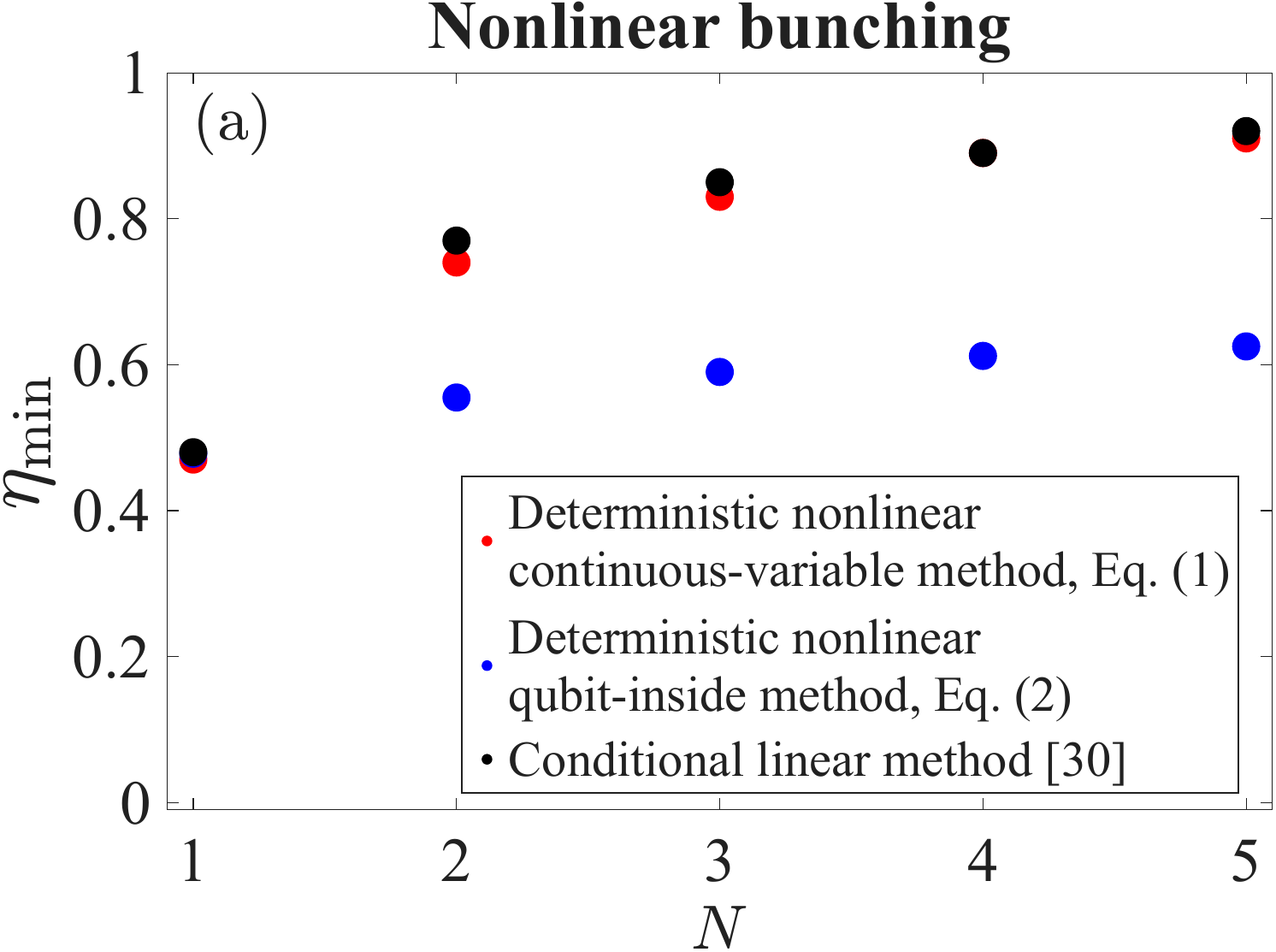}
    \includegraphics[width=0.49\columnwidth]{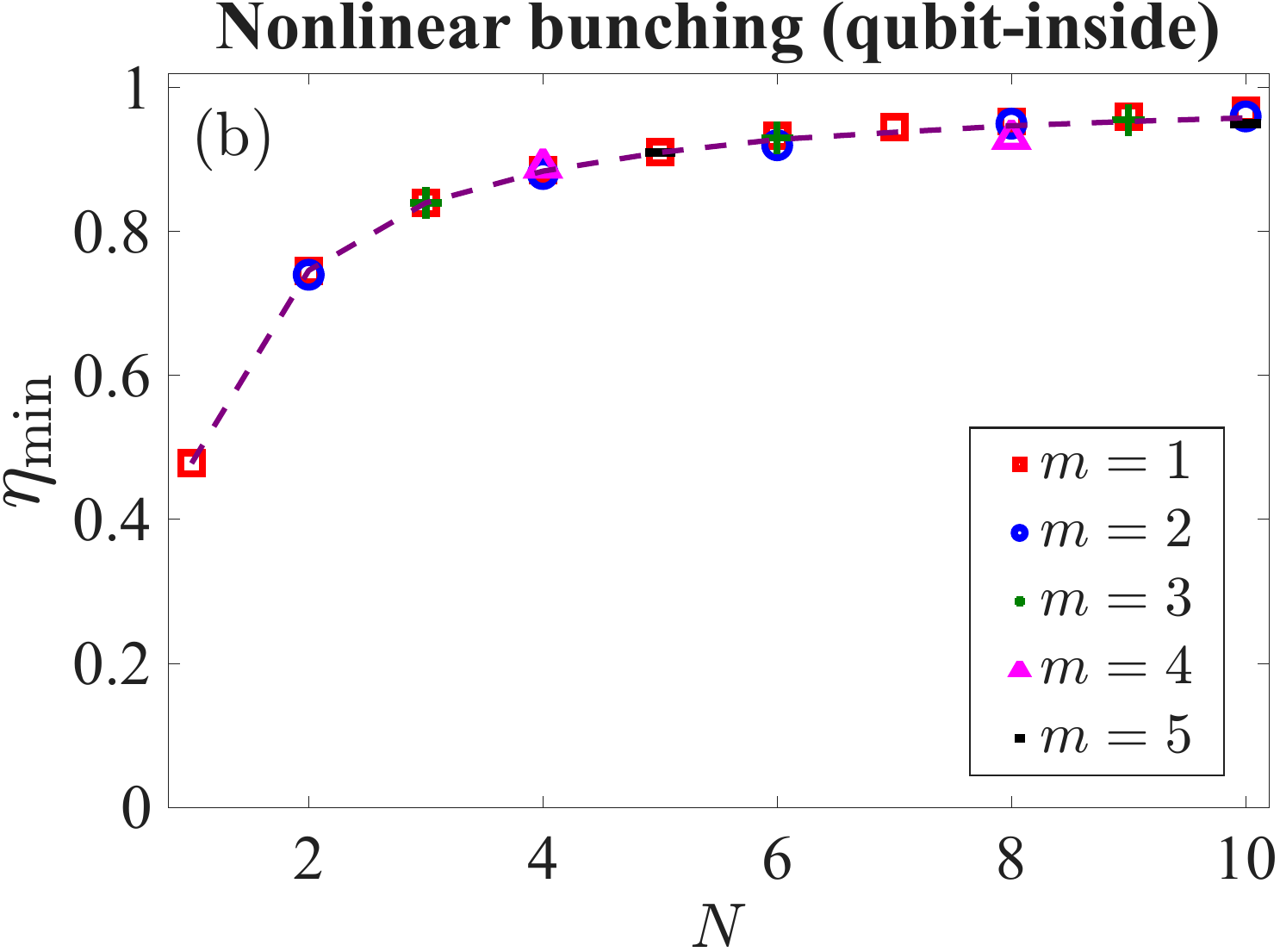}
    \caption{(a) The robustness of bunching from the nonlinear bunching interactions Eqs.~(\ref{MW}) or (\ref{MQ}) against loss $\eta$ in the input states for $m=N$. The minimum value of $\eta$ is selected when the maximum achievable $P_N$ falls below the thresholds for $n$-photon genuine QNG~\cite{podhora_quantum_2022}. The qubit-inside interaction is more robust than the entirely CV multimode interaction and than the conditional linear scheme~\cite{zapletal_multi-copy_2017}, further emphasising the advantages of nonlinearity. (b) Robustness to loss for the key interaction Eq.~(\ref{MQ}) (qubit-inside) for $m<N$, estimated by numerical simulation. The overlapping points demonstrate that nesting does not affect the minimum value of $\eta$ for which the QNG thresholds are still satisfied. The dashed line is the analytical prediction from graph theoretic analysis.}
    \label{m=Nloss}
\end{figure}

A basic feasibility test of the suppression of certain transitions is given by the introduction of lossy initial states as it introduces further possible transitions with differing total numbers of quanta. We model imperfect single boson states as statistical mixtures of the ground state (zero bosons) and the single boson state, $\ket{1}\bra{1}\rightarrow(1-\eta)\ket{0}\bra{0}+\eta\ket{1}\bra{1}$ with $\eta$ a loss parameter. This introduces a statistical mixing without introducing any higher order Fock states (although the advantages for qubit-inside interactions also hold for small thermal noise, see End Matter). Fig.~\ref{m=Nloss}~(a) shows that for $m=N$ the qubit-inside interactions are substantially more robust to loss than the multimode interactions as well as conditional linear methods~\cite{zapletal_multi-copy_2017}.

Advantageously the robustness to loss is independent of nonlinearity order $m$ for a given $N$, shown in Fig.~\ref{m=Nloss}~(b). That is, nesting does not reduce the capacity to bunch for pure single bosons or for those affected by initial loss. This suggests $m$ can be minimised without any reduction in $P_N$. That is, the $P_N$ resulting from the most costly $m=N$ case can be achieved by repeatedly using an internal resonant Jaynes-Cummings (JC) interaction, which corresponds to Eq.~(\ref{MQ}) with $m=1$. The saturability of the qubit, by suppressing certain transitions between fixed quanta states as in Fig.~\ref{SketchGraphical}, can very strongly enhance the bunching effect producing sufficient $n$-photon QNG of the output state. The reduction in the nonlinearity order $m$ afforded by the insensitivity of nesting to loss $\eta$ affecting the initial states is compensated by a concomitant increase in the time required for the nonlinear interactions (\ref{MQ}) to take place. The higher order nonlinearities typically have correspondingly larger transition frequencies and require fewer steps, or even just one, so that bunching occurs faster when more subsystems are involved, independent of the input loss. In realistic scenarios where decoherence plays a role this means there is a technical tradeoff between having more subsystems exposed to decoherence versus achieving the bunching at a faster rate. We show this scaling in the End Matter. Importantly, decreasing the nonlinearity order $m$ via nesting also reduces the number of simultaneously interacting subsystems, providing a further advantage despite the increased interaction times. 

{\it Discussion}---~Fig.~\ref{SketchGraphical} demonstrates graphically the importance of the saturation of the qubit on passive nonlinear bunching. We can formalise this, inspired by continuous time quantum walks~\cite{childs_example_2002,childs_universal_2009} on dynamic graphs~\cite{herrman_continuous-time_2019} and we do so in the End Matter. The key point is that a correspondence can be drawn between the Hamiltonians in Eqs.~(\ref{MW}-\ref{QM}) and the adjacency matrices of the graphs in Fig.~\ref{SketchGraphical}, up to graph weights. Interactions (\ref{MW}) and (\ref{MQ}) commute with the total boson number operator, partitioning the evolution so that distinct subgraphs representing subspaces with different total quanta numbers evolve separately and independently, preserving the initial probability of the state being on that subgraph. In the language of continuous time quantum walks, two-vertex path graphs (complete graphs $K_2$) enable perfect state transfer~\cite{christandl_perfect_2004,christandl_perfect_2005}, regardless of the graph weights. For pure states and $m=N$ the graphs formed by Eqs.~(\ref{MW},~\ref{MQ}) are $K_2$, enabling ideal bunching. For nesting, the saturation of the qubit ensures that the graph at each step is also $K_2$. The fully CV setting inevitably involves longer path graphs with weights determined by the Hamiltonian and the initial boson occupations, which cannot guarantee bunching as the graph weights cannot be independently modified~\cite{christandl_perfect_2005}. Further steps eventually take place on multiple disconnected subgraphs with different total quanta numbers, permanently preventing ideal bunching.

Similar reasoning applies to explain the identical robustness to initial loss for qubit-inside systems is identical across different nesting strategies. The derivation is detailed in the End Matter, but the key point again is that for both nesting and single-step protocols the evolution partitions along different subgraphs which are $K_2$. This allows tracking of the probability of ideal bunching. In both scenarios this probability is $\eta^N=(\eta^m)^{\frac{N}{m}}$, so that $P_N=\eta^N$ predicts the minimum $\eta$ for detection of $n$-photon QNG as in Fig.~\ref{m=Nloss}~(b).

Finally, a natural generalisation of single quanta bunching involves multiple, possibly different, numbers of quanta arriving in the same mode i.e. $\ket{k_i}^{\otimes p}$, $1\le i\le p<N$ instead of $\ket{1}^{\otimes N}$. (an example of fixed numbers $k_i=M$ is shown in Fig.~\ref{Sketch}). The qubit-inside nonlinearity with perfect state transfer on $K_2$ can be recovered at the cost of additional qubits. Each mode with $k_i$ bosons must interact sequentially via a basic JC interaction (Eq.~(\ref{QM}); $m=1$) with $k_i$ qubits, transferring an excitation to each one, thus there are $\sum_i^pk_i$ total qubits. Each of these qubits is then used in Eq.~(\ref{MQ}) to bunch all bosons. In the sketch for example, we have $MN$ total qubits required to enable bunching. However, here we see the full power of the qubit-inside strategy. As before, the number of qubits can always be reduced to 1 via nesting, at the cost of longer sequences of interactions.

{\it Conclusion}---~Our study of the bunching of oscillator energy quanta constitutes an operational way of categorising and understanding nonlinear bosonic interactions. It demonstrates how deterministic quantum processing can be accomplished by energy-conserving nonlinear couplings, which may have implications for energetically efficient versions of quantum technology~\cite{auffeves_quantum_2022}. Passive deterministic bunching of bosonic energy quanta requires nonlinearity. To do so efficiently in the face of imperfect initial states (single quanta with loss) further requires nonlinearity involving a saturable system, in this case a qubit. This is a proof-of-principle and operational example of the fundamental distinction between saturable and unsaturable nonlinear dynamics. For bunching, the symmetry given by energy conservation alongside the saturability of the qubit grants robustness to the process and a drastic lowering of both the number of subsystems and order of nonlinearity at the cost of longer bunching times.

This logic can in principle be extended to other tasks involving low quanta numbers beyond the primitive of large number states, such as error correcting codes~\cite{chuang_simple_1995,eczoo_binomial}, and without external driving or measurements, as many light-matter interactions can be described with graphical models~\cite{saugmann_fock-state-lattice_2023}. One may also consider superpositions in the Fock basis. The components of the superposition with different values of the total number operator evolve inside different subspaces (or on different subgraphs) so that the original probability amplitudes are preserved while the component Fock states may change drastically, potentially opening new methods to craft higher order finite superpositions from low order starting points, such as error correcting codes in the form of binomial, CLY and dual rail codes among many~\cite{eczoo_binomial}. Such low quanta (even thermal) regimes, in combination with nonlinear dynamics, have already been shown to give rise to surprising quantum phenomena~\cite{laha_entanglement_2024}.

\begin{acknowledgments}
We acknowledge the European Union’s HORIZON Research and Innovation Actions under Grant Agreement no. 101080173 (CLUSTEC), project CZ.02.01.010022\_0080004649 (QUEENTEC) of the EU and the Czech Ministry of Education, Youth and Sport (MEYS) and the project 8C24003, the MEYS CR, and EU under Grant Agreements no. 101017733 and 731473 (CLUSSTAR).
\end{acknowledgments}

\bibliographystyle{unsrt}
\bibliography{references}

\appendix

\section{Interaction time lengths}

\begin{figure}
    \centering
    
    \includegraphics[width=0.49\columnwidth]{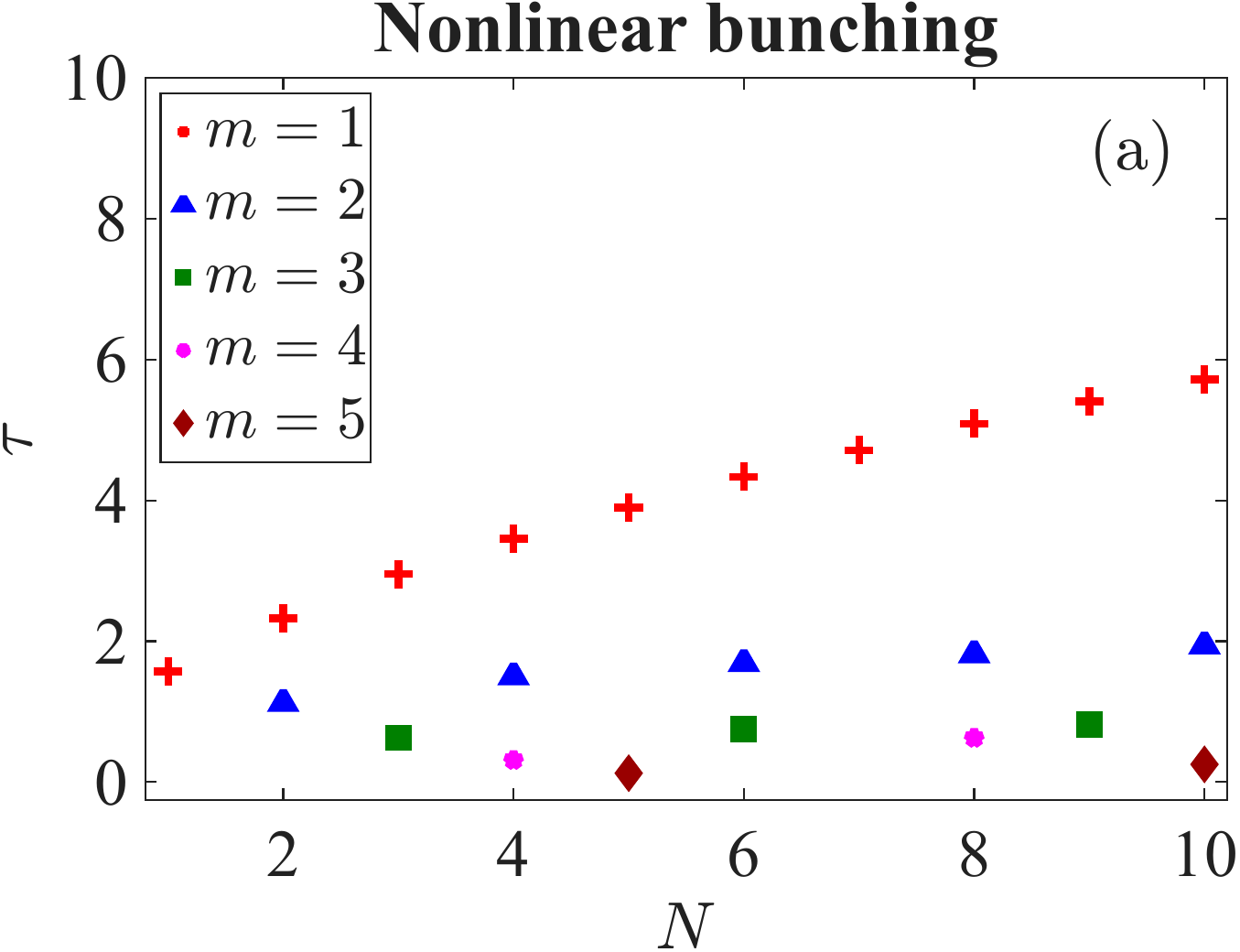}
        \includegraphics[width=0.49\columnwidth]{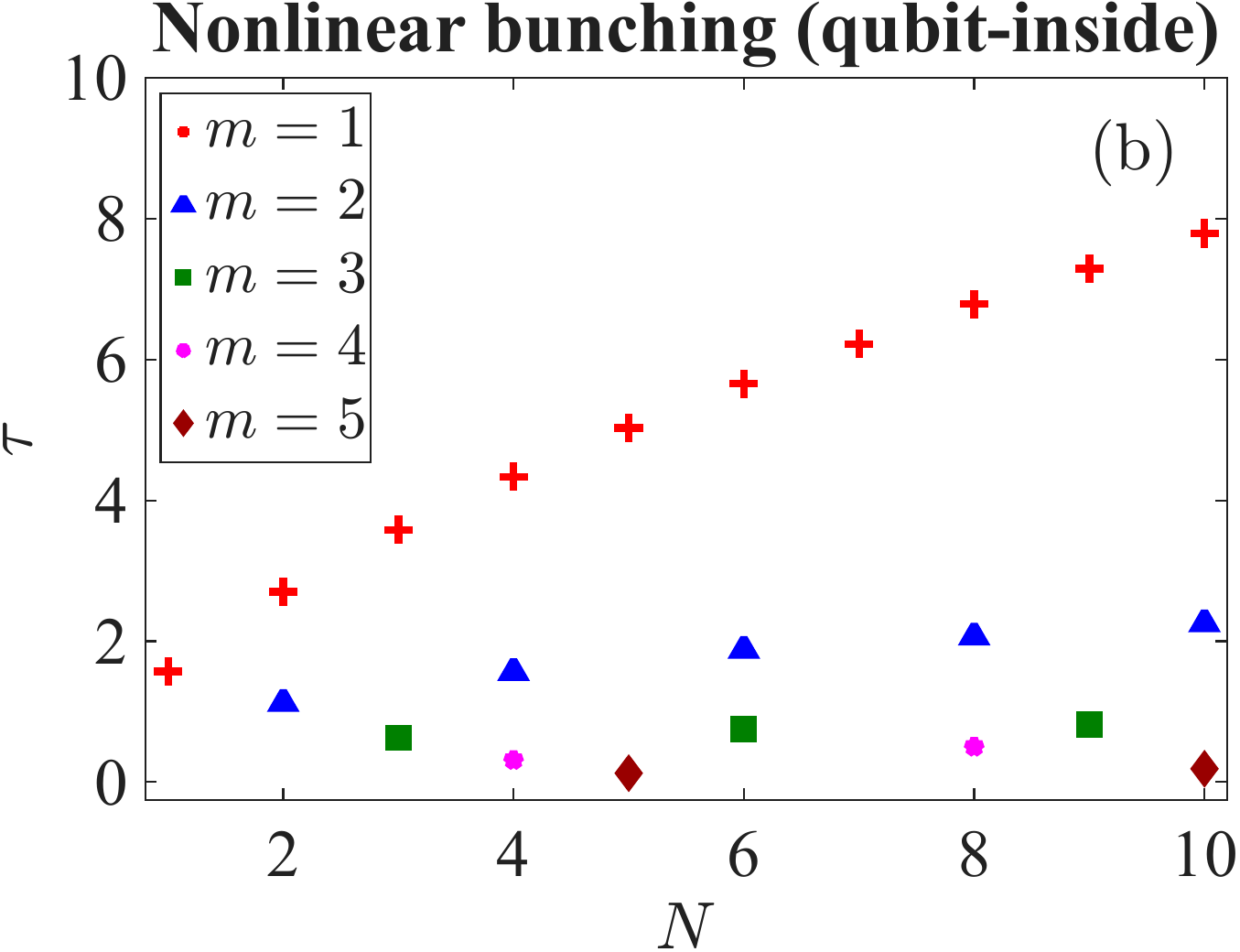}
        
    \caption{The total time $\tau$ to bunch $N$ bosons with the nesting scheme using (left) fully CV interactions, Eq.~(\ref{QM}), and (right) qubits, Eqs.~(\ref{MQ},~\ref{QM}). For qubit-inside interactions reducing $m$ to 1 increases the time required for the bunching to occur. Notably, CV interactions can be faster and do not require extra time (not shown) to transfer the bosons to qubits but, as shown in the main text, have substantially lower bunching probabilities.}
    \label{times}
\end{figure}

The qubit-inside interaction shows an independence of the index $m$ for bunching (Fig.~\ref{m=Nloss}) and thus the nesting strategy can be minimised to the $m=1$ case, a resonant Jaynes-Cummings type interaction. As noted in the main text, this suggests a substantial advantage in that the order of the nonlinearity and the number of subsystems can be substantially reduced. However it also maximises the number of nesting steps required to complete the bunching. This results in a tradeoff where the minimalisation results in longer interaction times with a consequently greater exposure to decoherence during the evolution. The scaling of these different $m$ as a function of $N$ is demonstrated in Fig.~\ref{times}. Due to the graph structure of the evolution (see next section), initial loss does not affect these times, only the probability of ideal bunching.

\section{Initial thermal noise}

\begin{figure}
    \centering
    \includegraphics[width=0.49\linewidth]{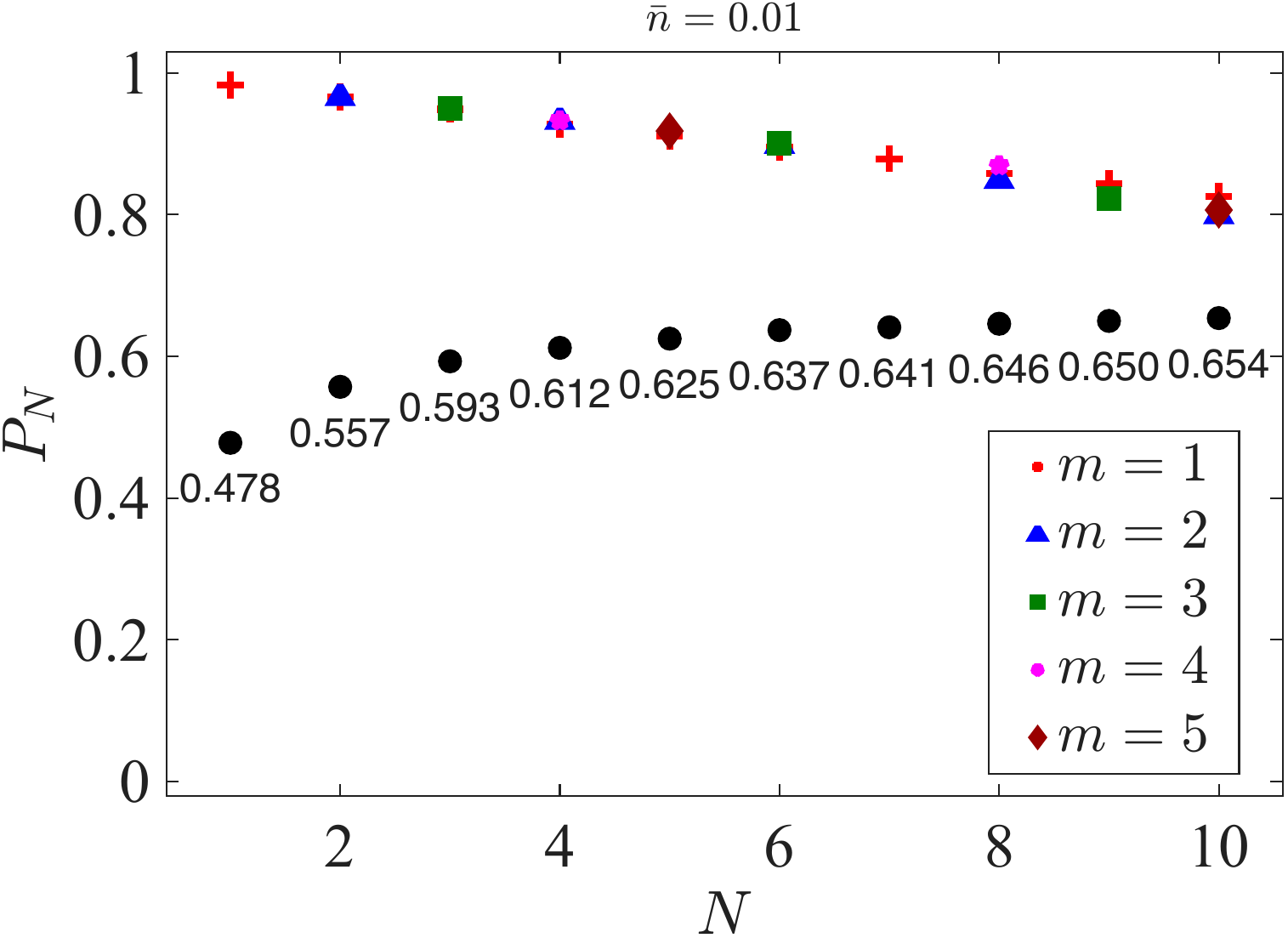}
    \includegraphics[width=0.49\linewidth]{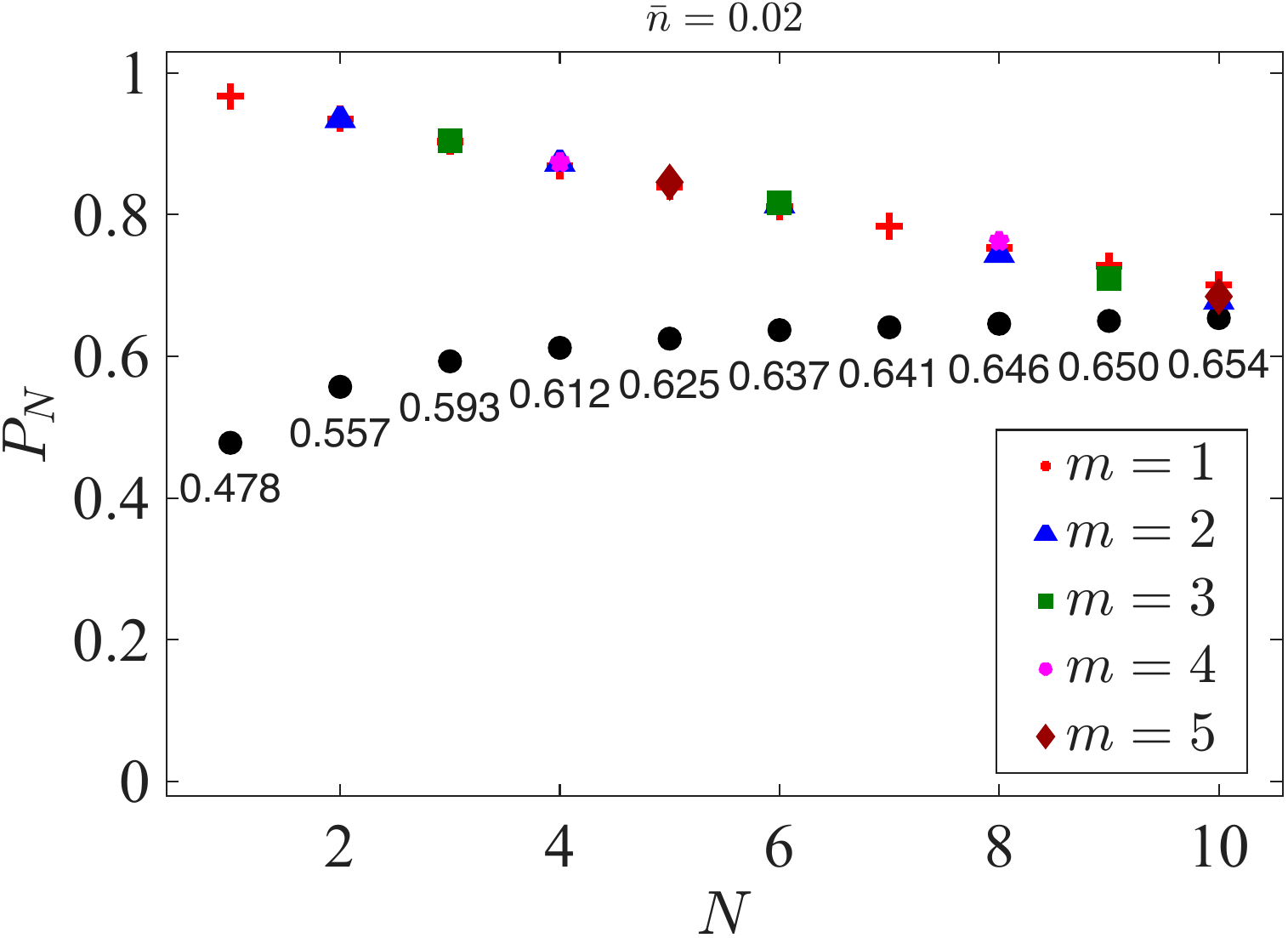}
    \caption{Nonlinear bunching using the qubits-inside nonlinearity based on interactions Eq.~(\ref{MQ}) for $m\le N$ for single bosons with added initial thermal noises $\nb=0.01,~0.02$
    The effectiveness of nesting is again advantageously independent of $m$ (all markers overlap at fixed $N$) and the sensitivity to noise for a greater target $N$ is similarly linear compared to an exponential decrease without qubits inside (see Fig~\ref{m<Npure}).}
    \label{Thermal}
\end{figure}

In the main text we discuss the effect of initial loss on the incoming bosons. Loss can only decrease the number of bosons, whereas noise adds bosons. We consider the following thermalisation map, $\mathcal{M}_{\nb}(\rho)=\int\frac{d^2\alpha}{\pi\nb}e^{-\frac{|\alpha|^2}{\nb}}D(\alpha)\rho D^\dagger(\alpha)$, where $\nb$ is the number of added thermal bosons. Fig.~\ref{Thermal} shows that the advantages of the qubit-inside strategy persist also in the case of small thermal noise.

\section{Bunching as continuous time quantum walks}

We fix some notation. The interactions (\ref{MW}) and (\ref{MQ}) were chosen with the physical motivation that no bosonic energy quanta are added or subtracted by the dynamics; they are energetically passive. This translates to the existence of a generalised total number operator $\mathcal{N}$ which is conserved by the dynamics: $[\mathcal{N},H]=0$. We have $\mathcal{N}_{\text{MM}}=\sum_{i=0}^ma_i^\dagger a_i$ and $\mathcal{N}_{\text{MQ}}=a_0^\dagger a_0+\sum_{i=1}^m\sigma_+^{(i)}\sigma_-^{(i)}$. Each CV mode $i$ has an eigenbasis ${\ket{k}_i}$ due to the Hermitian operator $a_i^\dagger a_i$ and each qubit has the eigenbasis $\{\ket{l}_i\}_{l\in\{0,1\}}$ associated with $\sigma_+^{(i)}\sigma_-^{(i)}$. The tensor product of the relevant eigenbases constructs a basis for the whole Hilbert space, where for $\mathcal{N}_{MM}$ we write $\{\bigotimes_i\ket{k}_i\equiv\ket{k_0,k_1,\dots,k_m}\}$ and for $\mathcal{N}_{MQ}$ we write $\{\ket{k}_0\otimes\bigotimes_i\ket{l}_i\equiv\ket{k_0,l_1,\dots,l_m}\}$. In each case the Hilbert space is partitioned into subspaces with $\sum_ik_i=\mathcal{N}_{\text{MM}}$ or $k_0+\sum_il_i=\mathcal{N}_{\text{MQ}}$. Since the pure initial states are always chosen with a fixed integer value of $\mathcal{N}$ the dynamics always remains inside its original subspace. 

To save repetition in what follows we focus on the case represented by $\mathcal{N}_\text{MM}$. The case for $\mathcal{N}_\text{MQ}$ proceeds similarly. For a given subspace labelled by a fixed integer $\mathcal{N}_\text{MM}$ the Schr\"{o}dinger equation is solved by the ansatz 
\begin{equation}
    \ket{\Psi}=\sum_{\substack{i=1\\\sum_jk_j=\mathcal{N}_\text{MM}}}c_i(t)\ket{k_0,k_1,\dots,k_m}_i\,.
\end{equation}
Then the Schr\"{o}dinger equation $i\partial_t\ket{\Psi}=H^{(m)}_{\text{MM}}\ket{\Psi}$ can be rewritten in terms of the probability amplitudes associated with the subspace,
\begin{equation}
    i\partial_t\bm{c}(t)=A\bm{c}(t)\,,
\end{equation}
where $\bm{c}$ collects the probability amplitudes $c_i(t)$ into a vector and $A=\bm{c}^\dagger H^{(m)}_{\text{MM}}\bm{c}$ is a symmetric matrix that can be identified with a weighted adjacency matrix of a graph~\cite{diestel_graph_2025}. The unitary operator that solves the Schr\"{o}dinger equation is then $e^{-iAt}$. The corresponding adjacency matrix $\mathcal{A}$ (i.e. with the weights set to 1) allows a useful visualisation of the possible transitions, as in Fig.~\ref{Sketch}. 

In the case $m=N$ both multimode and multiqubit interactions generate an adjacency matrix $A$ which represents a two-vertex path graph, or the complete graph $K_2$. There is only one possible transition and the system oscillates between the two states (vertices). However for $m<N$ (nesting) this equivalence is broken as at each step $m$ photons are added, changing the value of $\mathcal{N}$ and thus potentially changing the graph structure. In particular, for the multimode interaction the addition of $m$ single photons at the second step introduces new possible transitions. These typically evolve anharmonically which generally prevents ideal bunching in mode 0. Moreover when ideal bunching is not achieved at one step, the next addition of $m$ photons in the following step results in multiple values of $\mathcal{N}_\text{MM}$. These subspaces evolve independently and can be understood as the disconnected subgraphs in Fig.~\ref{SketchGraphical}. This is a result of the linearity of unitary evolution on mixed states, $\rho=\sum_ip_i\ket{i}\bra{i}\rightarrow U\rho U^\dagger=\sum_ip_iU\ket{i}\bra{i}U^\dagger$. If each component $\rho_i$ has a different value of $\mathcal{N}$ it evolves independently on its own subgraph, with $p_i$ preserved. A similar result holds for initial states that are superpositions of states with different $\mathcal{N}$. Since these subspaces never rejoin ideal bunching is now permanently prevented. In crucial contrast, the saturability of the multiqubit interaction ensures that the graph structure {\it does not change} after the first step. This means that at each step the system oscillates again between two states and ideal bunching can always be achieved.

The graph structure can also help explain why the robustness to initial loss for qubit-inside systems is identical across different nesting strategies (Fig.~\ref{m=Nloss}~(b)). For a given multiqubit interaction with index $m$, the initial state including the loss is the density operator
\begin{equation}
    [\eta\ket{1}\bra{1}+(1-\eta)\ket{0}\bra{0}]^{\otimes m}\,.
\end{equation}
The linearity of unitary evolution means the components of the mixture, each with different total quanta number, evolve on separate subgraphs preserving their initial probability. The component of this initial state contributing to successful bunching is $\eta^m\ket{1}\bra{1}^{\otimes m}$, where $\eta^m$ is a probability. As the only component with this number of quanta, the probability $\eta^m$ is preserved in the transition to the bunched state. Suppose first that $m=N$, so that bunching can occur in a single step. Then $P_N=\eta^N$ predicts the minimum $\eta$ for detection of $n$-photon QNG as in Fig.~\ref{m=Nloss}~(b). This also holds for nested schemes with $m\le N$. In a second step (Fig.~\ref{Sketch}~(d)) the state of the previous output mode will take the general form $\rho^{\prime}_1=\sum_{i=0}^mp_i\ket{i}\bra{i}$, where we can specify $p_m=\eta^m$. Then the initial state for the second step is
\begin{equation}
    \rho^{\prime}_1\otimes[\eta\ket{1}\bra{1}+(1-\eta)\ket{0}\bra{0}]^{\otimes m}\,,
\end{equation}
whose contributing component for successful bunching is $\eta^m\ket{m}\bra{m}\otimes\eta^m\ket{1}\bra{1}^{\otimes m}$, with probability $\eta^{2m}$. The pattern continues until the end of the nesting sequence where finally the exponent of $\eta$ is $N$, identical to the single step case where $m=N$. This explains why even in the presence of initial loss the capability to replace higher order nonlinearities with multiple instances of JC ($m=1$) interactions is still valid.

\end{document}